\documentstyle[aps]{revtex}
\newcommand{\va}[1]{\langle{#1}\rangle}                                  
\newcommand{\dpa}{\partial}                                              
\newcommand{\sto}[4]{#1\left(\tiny\begin{array}{c}                       
 {\mathbf #2}\\{\mathbf #3}\\{\mathbf #4}\end{array}\right)}             
\newcommand{\fin}[2]{\varphi^{\mathbf{#1}}_{\mathbf{#2}}}                
\newcommand{\fic}[2]{\varphi^{*\mathbf{#1}}_{\mathbf{#2}}}               
\newcommand{\scn}[2]{c^{\phantom{\dagger}}_{\mathstrut#1\mathbf{#2}}}    
\newcommand{\scc}[2]{c^{\dagger}_{\mathstrut#1\mathbf{#2}}}              
\newcommand{\ccn}[2]{C^{\phantom{\dagger}}_{\mathstrut#1\mathbf{#2}}}    
\newcommand{\ccc}[2]{C^{\dagger}_{\mathstrut#1\mathbf{#2}}}              
\newcommand{\tcn}[2]{\tilde{C}^{\phantom{\dagger}}_{\mathstrut#1\mathbf{#2}}}
\newcommand{\tcc}[2]{\tilde{C}^{\dagger}_{\mathstrut#1\mathbf{#2}}}      

\newcommand{\rpp}[2]{\rho^{\phantom{\dagger}}_{\mathstrut#1\mathbf{#2}}} 
\newcommand{\trp}[2]{\tilde{\rho}^{\phantom{\dagger}}_{\mathstrut#1\mathbf{#2}}}
\newcommand{\bbn}[1]{B^{\phantom{\dagger}}_{\mathstrut\mathbf{#1}}}      
\newcommand{\bbc}[1]{B^{\dagger}_{\mathstrut\mathbf{#1}}}                
\newcommand{\sbn}[1]{b^{\phantom{\dagger}}_{\mathstrut\mathbf{#1}}}      
\newcommand{\sbc}[1]{b^{\dagger}_{\mathstrut\mathbf{#1}}}                
\newcommand{\epp}[1]{\epsilon_{\mathbf{#1}}}                             

\begin{document}

\title{The Condition for the Onset of High Temperature Superconductivity.}

\author{I.~M.~Yurin and V.~B.~Kalinin}

\address{Institute of Physical Chemistry, Leninskiy prosp. 31,
         GSP-1, Moscow, 117915, Russia}
\maketitle

\begin{abstract}
In this work the long-wave limit of electron-electron interaction 
arising from the exchange of virtual phonons in an approximation close to
"jelly" model is considered. It is shown that the interaction through the
exchange of virtual phonons is actually not screened in contrast to the
Coulomb one; this just leads to instability relative to the formation of
pairs near the Fermi surface. The consequences of this approach are examined 
with respect to high-temperature superconducting materials that have recently
been synthesized. An approximate relationship connecting sound and Fermi
velocities for these materials is obtained.
\newline
PACS number(s): 74.20.-z, 74.20.Mn, 74.72.-h
\end{abstract}

\section{Introduction}

  A considerable number of fundamental works 
have been devoted to tackling the superconductivity problem 
and, in  particular, a high temperature superconductivity one (HTSC).
The opinion on the present state of the problem 
may be formed considering the works ~\cite{bib-1,bib-2,bib-3}.
Most relevant publication are to some extent associated 
with the BCS theory~\cite{bib-4}. Now, for metals enumerated in
the Periodic System of elements (PSE), the superconductivity theory
is considered to be more or less completed and the validity of the
BCS theory leaves no room for doubt on the part of the majority of
authors. At the same time in ~\cite{bib-5,bib-6,bib-7,bib-8} a conclusion
was drawn that the BCS theory may be not the only possible way to explain
the phenomenon of superconductivity (SC).

Outstanding results ~\cite{bib-9} obtained by Bednortz J.G. and Muller A.K.
in 1986 for YBa$_{2}$Cu$_{3}$O$_{7-\delta}$ caused the great sensation among
chemists, physisists and material researchers. Since that time the highest
value for T$_{s} $ at about 164K was observed in HgCa$_{2}$Ba$_{2}$Y$_{3}$Cu$_{8+\delta}$
ceramics. Rather high T$_{s} $ values later were observed in fullerenes  ~\cite{bib-10}
and their derivatives fullerides C$_{60} $M ~\cite{bib-11}. Some speculations show 
that in intercalates ~\cite{bib-12} and nanotubes ~\cite{bib-13} high T$_{s} $ values
may be observed. The idea of producing HTSC in simple composition compounds such as
Li$_{3}$P and Li$_{3}$N is presently in a research state ~\cite{bib-14,bib-15,bib-16}.

These experimental results led to the conclusion ~\cite{bib-3,bib-17} that the 
applicability of the BCS theory for HTSC is not so obvious as for
convenient SC and the theorists encountered many difficult problems
which are not yet solved. From our opinion the development of the SC theory was stopped
at the level determined by ~\cite{bib-2,bib-3}. It should be noted here,
that the electron-phonon (el-ph) interaction in the majority of publications,
referred to in ~\cite{bib-2,bib-3}, is considered for the short wave or optical phonons.
This becomes obvious from the fact, that the polaron problem in these works is solved
separately from the rest part of the problem, not taking into account Coulomb
electron-electron (el-el) interaction. It may be easily proved, that such kind of
renormalization procedures are unacceptable in the long-wave limit investigation
due to the singular behaviour of the initial el-ph interaction matrix elements.
From the other hand, the consistent accounting for Coulomb el-el interaction in the polaron
problem leads to the appearance of the "screening effect" for the el-ph interaction
matrix elements and subsequently reasonable expression (\ref{eq-15})
for the parameters of the operator transformations involved in the renormalization
procedure is derived.

 The object of the present work is to calculate 
the long-wave range component of el-el
interaction arising from the exchange of virtual phonons
accounting for the correlation effects in an electronic plasma.
An approximation close to "jelly' model ~\cite{bib-18} was assumed to
describe the el-ph interaction. Authors believe, that the proposed microscopic
model may help to predict new promising systems for observing HTSC.

\section{Description of the model}
 
  For simplification, let us consider the Hamiltonian $H_{tot}$
of an infinite (finite-size effects are ignored) monoatomic metal
with a parabolic conductivity zone as an initial one
(here and further we use units with $\hbar=1$):
\begin{eqnarray}
  H_{tot} 
   &=& H_0 + H_{pp} + H_{ee} + H_{ep}
  \ ,\label{eq-1} \\
  H_0
   &=& \sum_{\mu}
       \int \epp{p} \scc{\mu}{p} \scn{\mu}{p} d{\mathbf p}
     + \int \omega_{\mathbf{q}} \sbc{q} \sbn{q} d{\mathbf q}\ , 
    \qquad \left(\epp{p} = \frac{p^2}{2m}\ ,
                \ \ \omega_{\mathbf{q}} = s q \right)
  \label{eq-2} \\
  H_{pp} &=&
   A \int
    \left[
    2 \sbc{q} \sbn{q}
    - \sbc{q} \sbc{-q}                
    - \sbn{q} \sbn{-q}
    \right] \frac{d{\bf q}}{q}                
  \ ,\label{eq-3} \\
  H_{ee}
   &=& G_{ee}\sum_{\mu}\sum_{\nu}\int\!\!\!\int\!\!\!\int
        \scc{\mu}{p+q} \scc{\nu}{k-q} \scn{\nu}{k} \scn{\mu}{p}
          d{\mathbf p} d{\mathbf k} \frac{d{\mathbf q}}{q^2}
  \ ,\label{eq-4} \\
  H_{ep}
   &=& 
    - i G_{ep}\sum_{\mu}\int\!\!\!\int
       \scc{\mu}{p-q} \scn{\mu}{p}
        \left( \sbc{q} - \sbn{-q}                
         \right) 
          d{\mathbf p} \frac{d{\mathbf q}}{q^{3/2}}
  \ ,\label{eq-5}
\end{eqnarray} 
where $A=\pi\frac{e^2}{\epsilon}\frac{z^2}{Ms\Omega}$,
 $G_{ee}=\frac{e^2}{4\pi^2\epsilon}$,
 $G_{ep}=\frac{e^2}{\epsilon}\frac{z}{\sqrt{\pi Ms\Omega}}$,
 $\scc{\mu}{p}$ and $\scn{\mu}{p}$
 are operators of creation and annihilation of electrons with
 momentum $p$, respectively, $\mu$, $\nu$ are spin indices,
 $\sbc{q}$,
 $\sbn{q}$
 are operators of creation and annihilation 
 of longitudinally polarized phonons with momentum $q$, respectively,
 $s$, $\epsilon$ are sound velocity and dielectric permittivity
 of the valence skeleton, respectively, 
 $M$ is the mass of ion, 
 $m$ is the mass of zone electron, 
 $\Omega$ is the volume of an elementary cell,
 $z$ is the number of conductivity electrons per cell,
 and $i$ is an imaginary unit. 

 Eqs.(\ref{eq-1}-\ref{eq-5}) are derived 
under the following assumptions. 
It is possible to assume that the valence skeleton
creates average positively charged background 
with a density $\rho_0=ze/\Omega$.
The density deviations from the mean value 
are determined by a relative change in the crystal volume 
resulting from the deformation of the lattice, 
when a longitudinal sound wave propagates through it.
At the same time operators of creation and annihilation 
of longitudinally polarized phonons are associated 
with the local shift of ions ${\bf u}(r,t)$
via a relationship
\begin{eqnarray}
 {\bf u}_\alpha 
  &=& \frac{1}{4\pi^{3/2}}
       \int\sqrt{\frac{\Omega}{M\omega_{{\mathbf q}}}}
        e^{-i{\bf q}{\bf r}}
         {\bf e}_{\alpha q}
          \sbc{q} d{\bf q} + \mbox{H.c.}
 \label{eq-6}\\
 {\bf e}_{\alpha q}
  &=& \frac{{\bf q}_\alpha}{q}\ ,
  \quad \alpha = x, y, z\ .\nonumber
\end{eqnarray}
     The choice of a factor in Eq.(\ref{eq-6}) is determined 
by need to fulfill the relationship 
$E = \frac{Ms^2}{2\Omega} \int 
     \sum_{\alpha}
     \left(\frac{\dpa{\bf u}_\alpha}
                {\dpa{\bf r}_\alpha}\right)^2 d{\bf r}
   + \frac{M}{2\Omega} \int 
     \left(\frac{\dpa{\bf u}}{\dpa t}\right)^2 d{\bf r}
  = \int \omega_{{\mathbf q}} 
     \sbc{q}
      \sbn{q} d{\bf q}$, 
where $E$ is the energy of longitudinally polarized phonons.
Thus, the local charge density is determined 
by the expression
$\rho_v = \rho_0 
  \det^{-1}\left|\frac{\dpa({\bf r}+{\bf u})}{\dpa{\bf r}}\right|$.
Taking into account only the terms $\sim \mbox{div}({\bf u}(r))$
one can get for deviation of local density from its mean value 
$\rho=\rho_v-\rho_0$ 
\begin{equation}
 \rho 
 = - \frac{ze}{\Omega}
      \sum_{\alpha}
       \frac{\dpa{\bf u}_\alpha}{\dpa{\bf r}_\alpha}
 = i \frac{ze}{4\Omega\pi^{3/2}}
      \int\sqrt{\frac{\Omega q}{Ms}}
        e^{-i{\bf q}{\bf r}}
         \sbc{q} d{\bf q} + \mbox{H.c.}
  \label{eq-7}
\end{equation} 
The derived expression for $\rho$ determines the potential
due to a phonon subsystem: 
\begin{equation}
 V_\rho =
  i z \frac{e}{\epsilon}
   \frac1{\sqrt{\pi Ms\Omega}}
    \int \sbc{q} 
      e^{-i{\bf q}{\bf r}}
       \frac{d{\bf q}}{q^{3/2}} + \mbox{H.c.}
  \label{eq-8}
\end{equation} 
It is this potential that causes in the suggested model
the existence of an electron-phonon interaction~(\ref{eq-5})
and a specific self-action of the phonon subsystem~(\ref{eq-3}).

 It should be noted here that in PSE metals the division of electron states
between conductivity and valence ones encounters difficulties due to
hybridization effects when these states belong to the same
energy interval. Nevertheless in the majority of HTSC materials
with low concentration of electro-active effects these difficulties
do not exist and the model developed may be useful in consideration of
el-ph and phonon-phonon interactions.

 In Eqs.~(\ref{eq-1})-(\ref{eq-5}) a dielectric permittivity 
of the valence framework $\epsilon$ was introduced. 
We note here that, as will become apparent below, 
the static long-wave limit of this value is of current significance. 

  As regards the sound velocity in the valence skeleton
we will derive an expression (see Eq.~(\ref{eq-31})) relating $s$
to the experimentally observed sound velocity in metal.
Therefore, in our opinion, the problem of $s$ determination
does not involve serious difficulties.
Moreover, from Eq.(\ref{eq-31}) it becomes apparent
that the suggested model transforms into a so-called
"jelly" model as $s \to 0$ (see Eq.(\ref{eq-32})), 
that must be obvious, from our point of view, 
without relationship~(\ref{eq-31}), too. 

  Actually, as compared with the "jelly" model, 
Eqs.(\ref{eq-1})-(\ref{eq-5}) 
take into account the rigidity of the valence skeleton.
This is of current significance in considering, 
for example, doped semiconductors. 

  Finally, let us consider the problem of 
taking into consideration transversely polarized phonons.
It is believed that the interaction of these phonons with electrons 
should be considerably smaller than that 
of longitudinally polarized ones.  
This is associated with fact 
that the elementary volume in shear deformations 
changes only in the second order 
with respect to the vector of elementary shift ${\bf u}(r)$.
That is why in the framework of the proposed investigation 
their existence may be ignored,
considering the subsystem of transversely polarized phonons 
not interacting both conductivity electrons 
and longitudinally polarized phonons.

\section{Polaron problem in proposed model}

  The Hamiltonian of interaction~(\ref{eq-5})
contains nondiagonal terms with structures $c^{\dagger}cb^{\dagger}$
and $c^{\dagger}cb$, 
corresponding to transitions without conservation of energy.
Therefore Fock states of the system, 
constructed on the basis of operators $c^{\dagger}$ and $b^{\dagger}$,
turn out to be states with indefinite energy.
It is this circumstance that creates difficulties 
in investigating low-energy excitations of the system, 
which current interest is obvious 
in describing the system at low temperatures. 

  Let us consider a transformation by introducing new operators 
$\ccc{}{}$, $C$, $\bbc{}{}$, and $B$,
being typical in solving polaron problems:
\begin{eqnarray}
 \scc{\mu}{p}
 &=& \ccc{\mu}{p} 
 - \int \fin{p}{q} \ccc{\mu}{p-q} \bbc{q} d{\bf q}
 + \int \fic{p+q}{q} \ccc{\mu}{p+q} \bbn{q} d{\bf q}
 - \frac{1}{2}
   \int \fic{p}{q}\fin{p}{q} d{\bf q} \ccc{\mu}{p} 
 \nonumber\\ &&
 + \frac{1}{2} \int\!\!\!\int
    \left[\fic{p-q}{-q}\fin{k}{-q}
        - \fic{k+q}{q}\fin{p}{q}
    \right]
     \ccc{\mu}{p-q} \sum_{\nu} \ccc{\nu}{k+q} \ccn{\nu}{k}
      d{\bf k} d{\bf q}
 \nonumber\\ &&
 - \frac{1}{2} \int\!\!\!\int
    \left[\fic{p-k+q}{q}\fin{p}{k}
        + \fic{p+q}{q}\fin{p+q}{k}
    \right]
     \ccc{\mu}{p-k+q} \bbc{k} \bbn{q}
      d{\bf k} d{\bf q}
 \label{eq-9} \\ &&
 + \frac{1}{4} \int\!\!\!\int
    \left[\fin{p-k}{q}\fin{p}{k}
        + \fin{p-q}{k}\fin{p}{q}
    \right]
     \ccc{\mu}{p-k-q} \bbc{q} \bbc{k}
      d{\bf k} d{\bf q}
 \nonumber\\ &&
 + \frac{1}{4} \int\!\!\!\int
    \left[\fic{p+q}{q}\fic{p+k+q}{k}
        + \fic{p+k}{k}\fic{p+k+q}{q}
    \right]
     \ccc{\mu}{p+k+q} \bbn{q} \bbn{k}
      d{\bf k} d{\bf q}\ ;
 \nonumber\\
 \sbc{q}
 &=& \bbc{q} 
 + \int \fic{p}{q}
    \sum_{\mu} \ccc{\mu}{p} \ccn{\mu}{p-q}
      d{\bf p}
 \nonumber\\ &&
 + \frac{1}{2} \int\!\!\!\int
 \left[ \fic{p}{q} \fin{p+k-q}{k}
      - \fic{p+k}{q} \fin{p+k}{k}
 \right]
  \sum_{\mu} \ccc{\mu}{p} \ccn{\mu}{p+k-q} \bbc{k}
    d{\bf p} d{\bf k}
 \label{eq-10} \\ &&
 + \frac{1}{2} \int\!\!\!\int
    \left[ \fic{p-k}{q} \fic{p}{k} - \fic{p-q}{k} \fic{p}{q}
    \right]
     \sum_{\mu} \ccc{\mu}{p} \ccn{\mu}{p-k-q} \bbn{k}
      d{\bf p} d{\bf k}\ ,
 \nonumber
\end{eqnarray}
where $\varphi$ stands for small parameter.

  It is easy to see (Appendix 1) that this transformation 
does not break commutation relations between 
the creation and annihilation operators 
with an accuracy up to terms $\sim |\varphi|^3$. 

  The parameters $\fin{p}{q}$ are involved 
to satisfy a requirement that the Hamiltontan 
expressed in terms of operators $C^{\dagger}$, $C$,
$B^{\dagger}$, and $B$ should not contain 
nondiagonal terms corresponding to virtual transitions,
i.e., to transitions without conservation of energy. 

  The necessity of fulfilling this requirement for the terms 
with the structure $C^{\dagger}CB^{\dagger}$
and $C^{\dagger}CB$ leads to the following equation 
for $\fin{p}{q}$ 
\begin{equation}
  - i G_{ep} q^{-3/2} 
  + \left(\epp{p-q} - \epp{p} + \omega_{\mathbf q}
    \right)\fin{p}{q} 
  + 2\frac{A}{q}
    \left(\fin{p}{q} - \fic{p-q}{-q}
    \right) 
  = 0\ .
 \label{eq-11}
\end{equation}

  Taking into account filling of the conductivity zone
by electrons leads us, in a sense of Hartree--Fock approximation,
to consider the terms with structures 
$C^{\dagger}C^{\dagger}CCB^{\dagger}$ and
$C^{\dagger}C^{\dagger}CCB$ 
and to realize their linearization 
with respect to the occupation dencity $\rpp{\mu}{p}$.
The linearization procedure is based
on the relation
$$
 \va{\ccc{\mu}{p}\ccn{\nu}{k}} 
 = \delta_{\mu}^{\nu}
    \rpp{\mu}{p}
     \delta({\mathbf p-k})\ ,
$$
where $\va{\ldots}$ means vacuum expectation value (vacuum corresponds to the
ground state of the system),
$\delta_{\mu}^{\nu}$ is Kronecker symbol,
$\rpp{\mu}{p}$ is occupation density,
$\delta({\mathbf p-k})$ is Dirac $\delta$-function and is carried out
according to usual rules
$$
\va{C^{\dagger}_1 C^{\dagger}_2 C_3 C_4 B^{\dagger}(B)}_L     
=(\va{C^{\dagger}_1 C_4}C^{\dagger}_2 C_3 
-\va{C^{\dagger}_1 C_3}C^{\dagger}_2 C_4
+\va{C^{\dagger}_2 C_3}C^{\dagger}_1 C_4
-\va{C^{\dagger}_2 C_4}C^{\dagger}_1 C_3 ) B^{\dagger}(B) 
$$
where $\va{A}_L$ means the result of the linearization procedure for the
operator expression $ A $.

  Linearization results in reducing of the corresponding Hamiltonian expressions
to the forms $\rho C^{\dagger}CB^{\dagger}$ 
and $\rho C^{\dagger}CB$.
As a consequence, one obtains a possibility
to include them into the equation for $\fin{p}{q}$, 
which, with taking into account additional terms,
appears to be as follows
\begin{equation}
  - i G_{ep} q^{-3/2} 
  + \left(\epp{p-q} - \epp{p} + \omega_{\mathbf q}
    \right)\fin{p}{q} 
  + 2\frac{A}{q}
    \left(\fin{p}{q} - \fic{p-q}{-q}
    \right) 
  + 2G_{ee}q^{-2} \sum_{\nu} \int
    \left(\fin{k}{q} - \fin{k+q}{q}
    \right)
     \rpp{\nu}{k} d{\mathbf k}
  = 0\ .
 \label{eq-12}
\end{equation} 

  The process of including the terms of a higher order
can be continued, for example, by bilinearization of the
$C^{\dagger} C^{\dagger} C^{\dagger} CCCB^{\dagger}(B)$ terms
according to $\rpp{\nu}{p}$.
Therefore, it is necessary to introduce hierarchy of arising terms
in order to single out the main ones in the equation for 
$\fin{p}{q}$. 

  We will assume that
\begin{equation}
 \sum_{\nu} \int
  \left(\fin{k+q}{q} - \fin{k}{q}
  \right)
   \rpp{\nu}{k} d{\bf k}
 = - i J_{q} \sim q^{1/2}\ .
 \label{eq-13}
\end{equation}

  The analysis of the first terms of the Hamiltonian $H_{tot}$ 
in variables $C^{\dagger}$, $C$, $B^{\dagger}$, and $B$,
included in the equation for $\fin{p}{q}$,
shows that the $n$-linearization 
leads to the appearance of a factor $J_{q}^n$ 
in corresponding equation terms.  
So, it follows that $J_q$, or $q^{1/2}$,
in this sense plays the role of a small parameter.
This conclusion does not contradict the remark
made with regard to Eqs.(\ref{eq-9})-(\ref{eq-10}).
Then from Eq.~(\ref{eq-13}) it is easy to see 
that the following equation for $\fin{p}{q}$
accounts for the main terms at $q \to 0$ 
\begin{eqnarray}
  - i G_{ep} q^{-3/2} 
  + 2 i G_{ee} q^{-2} J_q 
  + \left(\epp{p-q} - \epp{p} + \omega_{\mathbf q}
    \right)\fin{p}{q} 
  + 2\frac{A}{q}
    \left(\fin{p}{q} - \fic{p-q}{-q}
    \right) && \nonumber \\
  - \frac32 G_{ep}q^{-3/2} J_q
    \left(\fin{p}{q} - \fic{p-q}{-q}
    \right)
  + G_{ee}q^{-2} J_q^2
    \left(\fin{p}{q} - \fic{p-q}{-q}
    \right)
  &=& 0\ .
 \label{eq-14}
\end{eqnarray}

  Derivation of solution for $\fin{p}{q}$ 
according to Eq.(\ref{eq-14}) is not a very complicated procedure.
Therefore, it seems to be advisable simply 
to present the expression for $\fin{p}{q}$,
whose validity may be checked by a direct substitution
into Eq.~(\ref{eq-14}):
\begin{equation}
 \fin{p}{q}
 = i G_{ep}
    \frac{q^{1/2}}{q^2+\lambda^2}
     \frac{\epp{p} - \epp{p-q} + \omega_{\mathbf q}}
          {S^2q^2  - \left(\epp{p-q}-\epp{p}\right)^2}
 \ .
 \label{eq-15}
\end{equation}
where $S^2=s^2+(zm/6M)V_F^2$,
 $V_F=K_F/m$, 
 $\lambda^2=(4e^2/\pi\epsilon)mK_F$,
 $K_F$ is Fermi wave vector, 
 $K_F^3=(3\pi^2z/\Omega)$.

  In this case, for $J_q$ we derive the following expression:
\begin{equation}
  J_q 
  = 8 \pi m K_F G_{ep}\frac{q^{1/2}}{q^2+\lambda^2}\ ,
  \label{eq-16}
\end{equation}
being in accord with initial assumption~(\ref{eq-13}).
In deriving Eqs.(\ref{eq-15})-(\ref{eq-16})
we assumed that
\begin{equation}
 \rpp{\mu}{p} 
  = \left\{
     \begin{array}{lcr}
      1 & \mbox{for} & p < K_F \\
      0 & \mbox{for} & p > K_F
     \end{array}
    \right.\ ,
 \label{eq-17}
\end{equation}
which well describes the case of a degenerated electron gas 
at temperatures $T \ll E_F/k_B$, $E_F=K_F^2/2m$, 
$k_B$ is the Boltzman constant.
Attention should also be paid to a relationship 
(see Eq.~(\ref{eq-14})) 
\begin{eqnarray}
 && 2\frac{A}{q}
 - \frac{3}{2} G_{ep} q^{-3/2} J_q
 + G_{ee} q^{-2} J_q^2 
 = D q\ ;
 \label{eq-18}\\
 && D
 = \frac{\pi^2 z^2}{4 m M K_F \Omega s}
 \ ,\nonumber
\end{eqnarray}
which is fulfilled at an accuracy up to terms $\sim q^2$.
We will return to it later when discussing 
the problems of renormalization of the sound velocity 
in the model under consideration.

\section{Influence of screening on the phonon spectrum}

  The interaction of phonons with the electrons 
of the conductivity zone 
leads to an effect of renormalization of sound velocity.
The point is that experimentally observable sound velocity in metal
substantially differs from sound velocity in the valence skeleton s,
which was directly inserted into Eq.~(\ref{eq-2}).
In order to prove this statement,
let us substitute Eqs.(\ref{eq-9})-(\ref{eq-10})
in view of Eq.(\ref{eq-15}) into Eqs.(\ref{eq-1})-(\ref{eq-5})
and single out terms with the structures 
$\bbc{}B$, $\bbc{}\bbc{}$, $BB$,
$\ccc{}{}C\bbc{}\bbc{}$,
$\ccc{}{}C\bbc{}\bbc{}$,
$\ccc{}{}CBB$,
$\ccc{}{}\ccc{}{}CC\bbc{}B$,
$\ccc{}{}\ccc{}{}CC\bbc{}\bbc{}$ and 
$\ccc{}{}\ccc{}{}CCBB$.
The terms containing $\ccc{}{}C$ are linearized
and terms containing $\ccc{}{}\ccc{}{}CC$ are bilinearized
according to a scheme described in Section 3.
After singling out the main terms at $q \to 0$
at an accuracy up to the terms $\sim M^{-1/2}$
($A$ and $s \sim M^{-1/2}$, 
 $G_{ep}$, $J_{q}$, and $\varphi \sim M^{-1/4}$),
it is possible to obtain the following expressions
for the energy of phonon system $H_{op}$:
\begin{eqnarray}
 H_{op} &=&
   \int\omega_{\mathbf q}
    \bbc{q}\bbn{q}
     d\mathbf q
 \label{eq-19} \\ &+&
   2A\int
    \bbc{q}\bbn{q}
     \frac{d\mathbf q}{q}
 - 2G_{ep}\int J_q
    \bbc{q}\bbn{q}
     \frac{d\mathbf q}{q^{3/2}}
 + 2G_{ee}\int J_q^2
    \bbc{q}\bbn{q}
     \frac{d\mathbf q}{q^2}
 \label{eq-20} \\ &-&
   A\int
    \bbc{q}\bbc{-q}
     \frac{d\mathbf q}{q}
 + G_{ep}\int J_q
    \bbc{q}\bbc{-q}
     \frac{d\mathbf q}{q^{3/2}}
 - G_{ee}\int J_q^2
    \bbc{q}\bbc{-q}
     \frac{d\mathbf q}{q^2}
 \label{eq-21} \\ &-&
   A\int
    \bbn{q}\bbn{-q}
     \frac{d\mathbf q}{q}
 + G_{ep}\int J_q
    \bbn{q}\bbn{-q}
     \frac{d\mathbf q}{q^{3/2}}
 - G_{ee}\int J_q^2
    \bbn{q}\bbn{-q}
     \frac{d\mathbf q}{q^2}
 \label{eq-22} \\ &+&
   \int\!\!\!\int
    \left[
     \left(\epp{k+q} - \epp{k} - \omega_{\mathbf q}
     \right)
      \left|\fin{k+q}{q}\right|^2
   - \left(\epp{k} - \epp{k-q} - \omega_{\mathbf q}
     \right) 
      \left|\fin{k}{q}\right|^2
    \right]
   \sum_{\nu}\rpp{\nu}{k}
    \bbc{q}\bbn{q}
     d{\mathbf q} d{\mathbf k}
 \label{eq-23} \\ &+&
   \int\!\!\!\int
    \left[ 
     \left( \epp{k} - \epp{k+q} \right) 
      \fin{k+q}{q}\fin{k}{-q}
       \sum_{\nu}\rpp{\nu}{k}
        \bbc{q}\bbc{-q}
   + \left(\epp{k} - \epp{k+q} \right) 
      \fic{k+q}{q}\fic{k}{-q}
       \sum_{\nu}\rpp{\nu}{k}
        \bbn{q}\bbn{-q}
    \right]
     d{\mathbf q} d{\mathbf k} 
 \label{eq-24}
\end{eqnarray}
From an easily verified relationship
\begin{equation}
 \frac{A}{q}
 - G_{ep} q^{-3/2} J_q
 + G_{ee} q^{-2} J_q^2 
 = 0\ ,
 \label{eq-28}
\end{equation}
which is fulfilled at an accuracy up to terms $\sim q^2$,
it is possible to conclude that 
the terms~(\ref{eq-20})-(\ref{eq-22})
do not contribute to the renormalization of sound velocity.
Thus, it is possible to state 
that the Hamiltonian $H_{pp}$ in Eq.(\ref{eq-3}),
describing the self-action of a phonon field,
turns out to be completely compensated,
when the screening action of the electrons 
of the conductivity zone has been accounted for. 
Presumably the appearance of a factor $A q^{-1}$ 
in the model under investigation 
may be considered to be the opening link 
in a chain $A q^{-1} \to G_{ep}q^{-3/2}J_q \to G_{ee}q^{-2}J_q^2$
leading to the compensation of the singularities 
induced by the factor $A q^{-1}$ in the Hamiltonian~(\ref{eq-1})
(see also~(\ref{eq-18})). 

  After the integration of the terms~(\ref{eq-23})-(\ref{eq-24})
and taking into account that
$S-s \sim M^{-1/2}$, $S^2-s^2 \sim M^{-1}$,
the problem of phonon spectrum determination
reduces to diagonalization of the Hamiltonian $H_{op}$:
\begin{eqnarray}
 H_{op} &=&
   \left(s+F\right)\int q \bbc{q} \bbn{q} d{\mathbf q}
 - \frac{F}{2}\int q 
    \left[\bbc{q} \bbc{-q} + \bbn{q}\bbn{-q}
    \right]
     d{\mathbf q} 
 \ ; \label{eq-29}\\
 F &=& \frac{z}{6mMs} K_F^2
 \ . \nonumber
\end{eqnarray}
This problem can be resolved
using Bogolubov (1947) transformation:
\begin{equation}
  \bbc{q} 
  = C_{\mathstrut\theta}
     \tilde{B}^{\dagger}_{\mathstrut\mathbf q}
  + S_{\mathstrut\theta}
    \tilde{B}^{\phantom{\dagger}}_{\mathstrut-\mathbf q}
 \ ;\qquad \ \ \
  \bbn{q}
  = C_{\mathstrut\theta} 
    \tilde{B}^{\phantom{\dagger}}_{\mathstrut\mathbf q}
  + S_{\mathstrut\theta}
    \tilde{B}^{\dagger}_{\mathstrut-\mathbf q}
 \ ,\label{eq-30}
\end{equation}
where
\begin{equation}
  C_{\mathstrut\theta} =
  \frac{\cos{\theta}}{\sqrt{\cos{2\theta}}}\ ;
  \qquad 
  S_{\mathstrut\theta} =
  \frac{\sin{\theta}}{\sqrt{\cos{2\theta}}}\ ;
  \qquad 
  (\sin{2\theta}=\frac{F}{s+F})\ .
 \nonumber
\end{equation}
Transition to the new operators 
$\tilde{B}^{\dagger}_{\mathstrut\mathbf q}$ and 
$\tilde{B}^{\phantom{\dagger}}_{\mathstrut\mathbf q}$
leads to the renormalization of sound velocity, given by the following expression
\begin{eqnarray}
 \tilde{S} &=& \sqrt{s^2+\frac{zm}{3M}V_F^2}
 \label{eq-31}
\end{eqnarray}
where $V_F=K_F/m$ is the Fermi velocity, 
and $\tilde{S}$ is the experimentally observed sound velocity in metal. 
At $s \to 0$ the sound velocity becomes equal 
to the Bohm--Steiver velocity $\tilde{S}_{\mathrm B\mathrm S}$
taken from the "jelly" model
\begin{equation}
 \tilde{S}_{\mathrm B\mathrm S}
 = V_F\sqrt{\frac{zm}{3M}}\ .
 \label{eq-32}
\end{equation}
On the other hand, the validity of Eq.(\ref{eq-31})
in the case of a dielectric material ($K_F=0$)
also casts no doubts.
The realization of Eq.(\ref{eq-31}) in two limits 
proves that the derived expression~(\ref{eq-31})
takes into account the main corrections in phonon spectrum 
associated with the filling of the conductivity zone with electrons.

\section{Correlation effects in the degenerated electron plasma}

  Transition to polaron operators of creation and annihilation
$\ccc{}{}$ and $C$ leads to the renormalization 
of the el-el interaction. 
This becomes evident when the terms~(\ref{eq-9})-~(\ref{eq-10})
are substituted into Eqs.(\ref{eq-1})-(\ref{eq-5}),
followed by singling out the terms with the structure 
$\ccc{}{}\ccc{}{}CC$.
If we limit ourselves to consideration of merely el-el interaction, 
the Hamiltonian of the system 
may be reduced to
\begin{equation}
  H_{\mathrm tot}
  = \sum_{\mu} \int \epp{p}
     \ccc{\mu}{p}\ccn{\mu}{p}
      d{\mathbf p}
  + \sum_{\mu}\sum_{\nu}\int\!\!\!\int\!\!\!\int 
     \sto{U}{p}{k}{q}
      \ccc{\mu}{p+q}\ccc{\nu}{k-q}
       \ccn{\nu}{k}\ccn{\mu}{p}
        d{\mathbf p} d{\mathbf k} d{\mathbf q}
 \label{eq-33}
\end{equation}
where
$$ \sto{U}{p}{k}{q} = G_{ee} q^{-2} + \delta V_{epe}\ , $$
\begin{equation}
 \delta V_{epe}
 = i\left[G_{ep}q^{-3/2}
           \left(\fin{k}{q} - \fic{k-q}{-q}
           \right)
        - G_{ee}q^{-2}J_q
           \left(\fin{p}{-q} - \fic{p+q}{q}
           \right)
    \right]\ .
 \label{eq-34}
\end{equation}

  Note that the main terms $\sim M^{-1/2}$ and $\sim q^{-2}$ at $q \to 0$
have been singled out in Eq.(\ref{eq-34}).
No renormalization of the electron kinetic energy (mass) 
due to the polaron effect has been done. 
This is associated with the fact 
that the kinetic energy of electron is assumed to be 
the highest of the values included in the present investigation. 
Therefore, its relative change cannot be substantial. 
As regards the term $\delta V_{epe}$,
in our formalism it is just the very term 
that is responsible for el-el interaction via 
virtual phonon exchange. 

  It is obvious that correlation effects
associated with the filling of the conductivity zone
should lead to a substantial change in the effective el-el interaction.
In connection with this, 
let us consider a transition to new, 
quasi-particle electron creation and annihilation operators
$\tcc{}{}$ and $\tilde{C}$: 
\begin{eqnarray}
  \ccc{\mu}{p}
  &=& \tcc{\mu}{p}
   - \sum_{\nu} \int 
      \sto{\theta_{\nu}^{\mu}}{p}{k}{q}
       \tcc{\mu}{p+q}\tcc{\nu}{k-q}\tcn{\nu}{k}
        d{\mathbf k} d{\mathbf q}
 \ ;\nonumber \\
  \ccn{\mu}{p}
  &=& \tcn{\mu}{p}
   - \sum_{\nu} \int 
      \sto{\theta_{\nu}^{*\mu}}{p}{k}{q}
       \tcc{\nu}{k}\tcn{\nu}{k-q}\tcn{\mu}{p+q}
        d{\mathbf k} d{\mathbf q}\ ,
 \label{eq-35}
\end{eqnarray}
where
\begin{eqnarray}
 \sto{\theta_{\nu}^{\mu}}{p}{k}{q}
  &=& \frac{1}{\epp{p}+\epp{k}-\epp{p+q}-\epp{k-q}}
      \sto{H_{\nu}^{\mu}}{p}{k}{q}
 \ ;\nonumber \\
 \sto{H_{\nu}^{\mu}}{p}{k}{q}
  &=& 2 \sto{\tilde{U}}{p}{k}{q}
      + \delta_{\nu}^{\mu}
       \left[2\sto{\tilde{V}}{p}{k}{q}
            - \sto{\tilde{U}}{p}{k}{q}
            - \sto{\tilde{U}}{p}{k}{k-p-q}
       \right]\ .
 \label{eq-36}
\end{eqnarray}

  For the terms $\tilde{U}$ and $\tilde{V}$,
without infringing on the general pattern of our arguments, 
we stipulate the fulfillment of the following conditions 
\begin{equation}
  \sto{\tilde{U}}{p}{k}{q}
  = \sto{\tilde{U}}{k}{p}{-q}
  \ ;\qquad
  \sto{\tilde{V}}{p}{k}{q}
  = \sto{\tilde{V}}{k}{p}{-q}
  \ ;\qquad
  \sto{\tilde{V}}{p}{k}{q}
  = - \sto{\tilde{V}}{p}{k}{k-p-q}
 \label{eq-37}
\end{equation}

  Let us note two remarks.
Firstly, the transformation~(\ref{eq-35})
does not break the commutation relations for operators 
$\tcc{}{}$ and $\tilde{C}$ 
at an accuracy up to the terms $\sim \theta^2$
(function $\theta$ now serves as a small parameter, see Appendix 2).
Secondly, it is transformation~(\ref{eq-35})
that diagonalizes in the first order of the perturbation theory in $\theta$
the Hamiltonian of the following form
(see Appendix 3) 
\begin{eqnarray}
 \tilde{H}_{\mathrm tot}
 &=& \sum_{\mu} \int \epp{p}
     \ccc{\mu}{p}\ccn{\mu}{p}
      d{\mathbf p}
  + \tilde{H}_{ee}
  \ ; \label{eq-38} \\
 \tilde{H}_{ee}
 &=& \sum_{\mu}\sum_{\nu}\int\!\!\!\int\!\!\!\int 
     \left[\sto{\tilde{U}}{p}{k}{q}
         + \sto{\tilde{V}}{p}{k}{q}\delta_{\nu}^{\mu}
     \right]
      \ccc{\mu}{p+q}\ccc{\nu}{k-q}
       \ccn{\nu}{k}\ccn{\mu}{p}
        d{\mathbf p} d{\mathbf k} d{\mathbf q}
 \label{eq-39}\ ,
\end{eqnarray}
It is easy to see that the condition~(\ref{eq-37})
comes from symmetries of~(\ref{eq-39}) under permutations,
while the Hermiticity requirement dictates 
the following relation: 
\begin{equation}
  \sto{\tilde{U}}{p}{k}{q}
  = \sto{\tilde{U}^{*}}{p+q}{k-q}{-q}
  \ ;\qquad
  \sto{\tilde{V}}{p}{k}{q}
  = \sto{\tilde{V}^{*}}{p+q}{k-q}{-q}
  \ ;\label{eq-40}
\end{equation}
and finally the symmetry of equations of motion 
under time reverse $t \to -t$ gives the following relations: 
\begin{equation}
  \sto{\tilde{U}}{p}{k}{q}
  = \sto{\tilde{U}}{-p}{-k}{-q}
  \ ;\qquad
  \sto{\tilde{V}}{p}{k}{q}
  = \sto{\tilde{V}}{-p}{-k}{-q}
  \ .\nonumber
\end{equation}

  It is obvious that the potentials $\tilde{U}$ and $\tilde{V}$
after introduction the self-consistent procedure 
will play the role of an effective el-el interaction potential,
accounting for the correlation effects of electron plasma.
The procedure for determination of $\tilde{U}$ and $\tilde{V}$ 
matrix elements should account for,
as in the case of the determination of $\fin{p}{q}$,
finite filling of the conductivity zone. 
For this purpose, similarly to what has been done in Section 3,
let us express the Hamiltonian~(\ref{eq-33})
in terms of operators $\tcc{}{}$ and $\tilde{C}$
using Eqs.(\ref{eq-36}). 

  The substitution results in the appearance of the terms
with the structure $\tcc{}{}\tcc{}{}\tcc{}{}\tilde{C}\tilde{C}\tilde{C}$
in the Hamiltonian $\tilde{H}_{\mathrm{tot}}$.
These terms may be linearized with respect to occupation density
$\trp{\mu}{p}$,
producing in this way the structure $\tilde{\rho}\tcc{}{}\tcc{}{}\tilde{C}\tilde{C}$.
This enables us to include these terms 
in the self-consistency equation for $\tilde{U}$ and $\tilde{V}$
of the following form
\begin{eqnarray}
  \sto{\tilde{U}}{p}{k}{q}
  + \delta_{\nu}^{\mu}
     \sto{\tilde{V}}{p}{k}{q}
 &=& \sto{U}{p}{k}{q}
  -  \sum_{\xi} \int
      \sto{U}{t}{k}{q}
      \sto{\theta_{\mu}^{\xi}}{t+q}{p}{-q}
       d{\mathbf t}
  -  \sum_{\xi} \int
      \sto{U}{p}{t}{q}
      \sto{\theta_{\nu}^{\xi}}{t-q}{k}{q}
       d{\mathbf t}
  \nonumber \\
 &-& \sum_{\xi} \int
      \sto{U}{p}{t+q}{q}
      \sto{\theta_{\nu}^{*\xi}}{t+q}{k-q}{-q}
       d{\mathbf t}
  -  \sum_{\xi} \int
      \sto{U}{t-q}{k}{q}
      \sto{\theta_{\mu}^{*\xi}}{t-q}{p+q}{q}
       d{\mathbf t}
  \nonumber \\
 &+& \int 
      \sto{U}{t}{k}{q}
      \sto{\theta_{\mu}^{\mu}}{t+q}{p}{p-t}
       d{\mathbf t}
  +  \int
      \sto{U}{p}{t}{q}
      \sto{\theta_{\nu}^{\nu}}{t-q}{k}{k-t}
       d{\mathbf t}
 \label{eq-41} \\
 &+& \int
      \sto{U}{p}{t+q}{q}
      \sto{\theta_{\nu}^{*\nu}}{t+q}{k-q}{k-q-t}
       d{\mathbf t}
  +  \int
      \sto{U}{t-q}{k}{q}
      \sto{\theta_{\mu}^{*\mu}}{t-q}{p+q}{p+q-t}
       d{\mathbf t}\ , \nonumber
\end{eqnarray}
the integration in Eq.~(\ref{eq-41}) being carried out 
in the region $t<K_F$,
it means that the case of low temperatures is considered. 
Note, that in the rhs of Eq.~(\ref{eq-41})
only the terms $\sim \theta$ and $\sim q^{-3}$
are presented. 

  It is easy to see from Eq.(\ref{eq-41})
that in order the term $\sto{\tilde{U}}{p}{k}{q}$ 
could satisfy conditions~(\ref{eq-37}) and~(\ref{eq-40}),
the term $\sto{{U}}{p}{k}{q}$ must also satisfy 
analogous relations: 
\begin{equation}
  \sto{U}{p}{k}{q}
  = \sto{U}{k}{p}{-q}
  \ ;\qquad
  \sto{U}{p}{k}{q}
  = \sto{U^{*}}{p+q}{k-q}{-q}
  \ ;\qquad \ldots \label{eq-42}
\end{equation}

  In connection with requirements of Eq.(\ref{eq-42}),
symmetrization of the term $U$ leads to 
the following expression for it: 
\begin{equation}
  \sto{U}{p}{k}{q}
  = \frac{G_{ee}}{q^2}
  + \frac{G_{ep}^2}{q^2+\lambda^2}
    \frac{ms}{2qS}
    \left[
     \pm\frac{1}{2{\mathbf pq}+q^2\mp 2mSq}
     \pm\frac{1}{2{\mathbf kq}-q^2\mp 2mSq}
    \right]
  \ , \label{eq-43}
\end{equation}
and in this case the Hamiltonian~(\ref{eq-33}) obviously is not changed.

  Although it would be preferable to solve Eq.(\ref{eq-41}) numerically,
from the point of view of the achievable accuracy of solution 
it seems to be of particular interest to obtain 
even a rough approximate solution without using computer technique:
such an approximate solution enables one 
to predict the character of limit relationships
in exact solution. 

  The essential point in the proposed solution scheme 
is the following approximate estimation of integrals 
inside the spheres with radius $K_F$:
\begin{equation}
 \int_{t<K_F} F({\mathbf t})S({\mathbf t})d{\mathbf t}
 \approx 
 F(0) \int_{t<K_F} S({\mathbf t})d{\mathbf t}
 \ ,\label{eq-44}
\end{equation} 
where $F(t)$ is regular, and $S(t)$ is the singular functions
of variable $t$.
It is obvious that this approximate relation is well performed 
when the character scale of change for $F$
turns out to be greater than $K_F$. 

  We will assume that the sought functions 
$\sto{\tilde{U}}{p}{k}{q}$ and
$\sto{\tilde{V}}{p}{k}{q}$ have the form 
\begin{equation}
  \sto{\tilde{U}}{p}{k}{q}
  = \sto{F_1}{p}{k}{q} S_1(q)
  \ ;\qquad
  \sto{\tilde{V}}{p}{k}{q}
  = \sto{F_2}{p}{k}{q} S_2(q)
  \ ;\label{eq-45}
\end{equation}
where $F_1$ and $F_2$ are regular,
whereas $S_1$ and $S_2$ are singular functions of their parameters.
Then only those terms should be left in Eq.(\ref{eq-41}),
in which the terms $S_l(q)$ can be put outside the integral,
and with the help of Eq.~(\ref{eq-44})
it is possible to obtain the following approximate equations
\begin{eqnarray}
 \sto{\tilde{V}}{p}{k}{q}
 &=& 0 
 \ ;\label{eq-46}\\
 \sto{\tilde{U}}{p}{k}{q}
 &=& \sto{U}{p}{k}{q} \nonumber\\
 &+& 
   4\sto{\tilde{U}}{p+q}{-q}{-q}
     \left\{
     \vphantom{\frac{G_{ep}^2}{2q^2}}
      \frac{mG_{ee}}{q^3}
      I\left[\frac{({\mathbf p+q}){\mathbf q}}{q} 
       \right]
    \right. \nonumber\\
 & &\qquad~~~~~~~~~~~
     + \frac{G_{ep}^2}{2q^2}\frac{m^2s/S}{q^2+\lambda^2}
        \left[\frac{1}{2{\mathbf kq} - q^2 - 2mSq}
            - \frac{1}{2{\mathbf kq} - q^2 + 2mSq}
        \right]
      I\left[\frac{({\mathbf p+q}){\mathbf q}}{q}\right]
 \nonumber \\
 & &\qquad~~~~~~~~~~~
     + \frac{G_{ep}^2}{2q^2}\frac{m^2s/S}{q^2+\lambda^2}
        \left[\frac{1}{2{\mathbf pq} + q^2 - 2mSq}
        \right]
        \left(I\left[\frac{({\mathbf p+q}){\mathbf q}}{q}
               \right]
            - I\left[mS+\frac{q}{2}
               \right]
        \right)
 \nonumber \\
 & &\qquad~~~~~~~~~~~\left.
     - \frac{G_{ep}^2}{2q^2}\frac{m^2s/S}{q^2+\lambda^2}
        \left[\frac{1}{2{\mathbf pq} + q^2 + 2mSq}
        \right]
        \left(I\left[\frac{({\mathbf p+q}){\mathbf q}}{q}
               \right]
            + I\left[mS-\frac{q}{2}
               \right]
        \right)
     \right\}
 \nonumber\\
 &-& 
   4\sto{\tilde{U}}{p}{q}{q}
     \left\{
     \vphantom{\frac{G_{ep}^2}{2q^2}}
      \frac{mG_{ee}}{q^3}
      I\left[\frac{{\mathbf pq}}{q} 
       \right]
    \right. \nonumber\\
 & &\qquad~~~~~~~~
     + \frac{G_{ep}^2}{2q^2}\frac{m^2s/S}{q^2+\lambda^2}
        \left[\frac{1}{2{\mathbf kq} - q^2 - 2mSq}
            - \frac{1}{2{\mathbf kq} - q^2 + 2mSq}
        \right]
      I\left[\frac{{\mathbf pq}}{q}\right]
 \nonumber \\
 & &\qquad~~~~~~~~
     + \frac{G_{ep}^2}{2q^2}\frac{m^2s/S}{q^2+\lambda^2}
        \left[\frac{1}{2{\mathbf pq} + q^2 - 2mSq}
        \right]
        \left(I\left[\frac{{\mathbf pq}}{q}
               \right]
            - I\left[mS-\frac{q}{2}
               \right]
        \right)
 \nonumber \\
 & &\qquad~~~~~~~~\left.
     - \frac{G_{ep}^2}{2q^2}\frac{m^2s/S}{q^2+\lambda^2}
        \left[\frac{1}{2{\mathbf pq} + q^2 + 2mSq}
        \right]
        \left(I\left[\frac{{\mathbf pq}}{q}
               \right]
            + I\left[mS+\frac{q}{2}
               \right]
        \right)
     \right\}
 \nonumber \\
 &+& 
   4\sto{\tilde{U}}{k}{-q}{-q}
     \left\{
     \vphantom{\frac{G_{ep}^2}{2q^2}}
      \frac{mG_{ee}}{q^3}
      I\left[\frac{{\mathbf kq}}{q} 
       \right]
    \right. \nonumber\\
 & &\qquad~~~~~~~~~
     + \frac{G_{ep}^2}{2q^2}\frac{m^2s/S}{q^2+\lambda^2}
        \left[\frac{1}{2{\mathbf kq} - q^2 - 2mSq}
        \right]
        \left(I\left[\frac{{\mathbf kq}}{q}
               \right]
            - I\left[mS+\frac{q}{2}
               \right]
        \right)
 \nonumber \\
 & &\qquad~~~~~~~~~
     - \frac{G_{ep}^2}{2q^2}\frac{m^2s/S}{q^2+\lambda^2}
        \left[\frac{1}{2{\mathbf kq} - q^2 + 2mSq}
        \right]
        \left(I\left[\frac{{\mathbf kq}}{q}
               \right]
            + I\left[mS-\frac{q}{2}
               \right]
        \right)
 \nonumber \\
 & &\qquad~~~~~~~~~\left.
     + \frac{G_{ep}^2}{2q^2}\frac{m^2s/S}{q^2+\lambda^2}
        \left[\frac{1}{2{\mathbf pq} + q^2 - 2mSq}
            - \frac{1}{2{\mathbf pq} + q^2 + 2mSq}
        \right]
      I\left[\frac{{\mathbf kq}}{q}\right]
     \right\}
 \nonumber \\
 &-& 
   4\sto{\tilde{U}}{k-q}{q}{q}
     \left\{
     \vphantom{\frac{G_{ep}^2}{2q^2}}
      \frac{mG_{ee}}{q^3}
      I\left[\frac{({\mathbf k-q}){\mathbf q}}{q} 
       \right]
    \right. \nonumber\\
 & &\qquad~~~~~~~~~~~
     + \frac{G_{ep}^2}{2q^2}\frac{m^2s/S}{q^2+\lambda^2}
        \left[\frac{1}{2{\mathbf kq} - q^2 - 2mSq}
        \right]
        \left(I\left[\frac{({\mathbf k-q}){\mathbf q}}{q}
               \right]
            - I\left[mS-\frac{q}{2}
               \right]
        \right)
 \nonumber \\
 & &\qquad~~~~~~~~~~~
     - \frac{G_{ep}^2}{2q^2}\frac{m^2s/S}{q^2+\lambda^2}
        \left[\frac{1}{2{\mathbf kq} - q^2 + 2mSq}
        \right]
        \left(I\left[\frac{({\mathbf k-q}){\mathbf q}}{q}
               \right]
            + I\left[mS+\frac{q}{2}
               \right]
        \right)
 \nonumber \\
 & &\qquad~~~~~~~~~~~\left.
     + \frac{G_{ep}^2}{2q^2}\frac{m^2s/S}{q^2+\lambda^2}
        \left[\frac{1}{2{\mathbf pq} + q^2 - 2mSq}
            - \frac{1}{2{\mathbf pq} + q^2 + 2mSq}
        \right]
      I\left[\frac{({\mathbf k-q}){\mathbf q}}{q}\right]
     \right\}\ ,
 \label{eq-47}
\end{eqnarray}
where
\begin{equation}
 I[a] = q \int_{t<K_F}\frac{d{\mathbf t}}{{\mathbf tq}-aq}
      = \pi K_F^2 \ln\left|\frac{K_F-a}{K_F+a}\right|
      - 2\pi a K_F\ .
 \label{eq-48}
\end{equation}
Note that at $a\ll K_F$ we have $I[a] \approx -4\pi a K_F$.

  The self-consistency Eq.(\ref{eq-47}) can be related
to equation of recurrent type. 
It is solved according to the scheme 
$\sto{\tilde{U}}{-q}{q}{q} \to
 \sto{\tilde{U}}{p}{q}{q} \to \sto{\tilde{U}}{p}{k}{q}$.
For the term $\sto{\tilde{U}}{-q}{q}{q}$
it is possible to derive the following expression:
\begin{equation}
 \sto{\tilde{U}}{-q}{q}{q}
 = \sto{U}{-q}{q}{q}
 + 8\sto{\tilde{U}}{-q}{q}{q}
     \left[
       \frac{mG_{ee}}{q^3}
     + \frac{G_{ep}^2}{2q^2}\frac{m^2s/S}{q^2+\lambda^2}
        \left(\frac1{q^2 - 2mSq} - \frac1{q^2 + 2mSq}
        \right)
     \right] I[q]
\end{equation}
and, thus, we get
\begin{equation}
 \sto{\tilde{U}}{-q}{q}{q}
 = \frac{G_{ee}}{q^2+2\lambda^2}
 + \frac{zm}{3M}K_F^2\lambda^{-2}\frac{G_{ee}}{q^2-\chi_1}
 \label{eq-49}
\end{equation}
where $\chi=4\left(m^2\tilde{S}^2-(zm/3M)K_F^2\right)$
(see Eqs.~(\ref{eq-15}),~(\ref{eq-31})),
$\chi_1=\left(1+(zm/3M)(K_F^2/\lambda^2)\right)\chi$, 
and one has $\chi_1-\chi\ll\chi$. 

  The direct substitutions into Eq.~(\ref{eq-47})
enables one to derive 
$\sto{\tilde{U}}{p+q}{-q}{-q} =
 \sto{\tilde{U}}{p}{q}{q}$.
Thus, we obtain an equation for 
$\sto{\tilde{U}}{p}{q}{q}$:
\begin{eqnarray}
 \sto{\tilde{U}}{p}{q}{q}
 &=& \sto{U}{p}{q}{q} \nonumber\\
 &+& 
   4\sto{\tilde{U}}{p}{q}{q}
     \left\{
     \vphantom{\frac{G_{ep}^2}{2q^2}}
      \frac{mG_{ee}}{q^3}
      I\left[\frac{({\mathbf p+q}){\mathbf q}}{q} 
       \right]
    \right. \nonumber\\
 & &\qquad~~~~~~~~~~~
    + \frac{G_{ep}^2}{2q^2}\frac{m^2s/S}{q^2+\lambda^2}
        \left[\frac1{q^2 - 2mSq} - \frac1{q^2 + 2mSq}
        \right]
      I\left[\frac{({\mathbf p+q}){\mathbf q}}{q}\right]
 \nonumber \\
 & &\qquad~~~~~~~~~~~
     + \frac{G_{ep}^2}{2q^2}\frac{m^2s/S}{q^2+\lambda^2}
        \left[\frac{1}{2{\mathbf pq} + q^2 - 2mSq}
        \right]
        \left(I\left[\frac{({\mathbf p+q}){\mathbf q}}{q}
               \right]
            - I\left[mS+\frac{q}{2}
               \right]
        \right)
 \nonumber \\
 & &\qquad~~~~~~~~~~~\left.
     - \frac{G_{ep}^2}{2q^2}\frac{m^2s/S}{q^2+\lambda^2}
        \left[\frac{1}{2{\mathbf pq} + q^2 + 2mSq}
        \right]
        \left(I\left[\frac{({\mathbf p+q}){\mathbf q}}{q}
               \right]
            + I\left[mS-\frac{q}{2}
               \right]
        \right)
     \right\}
 \nonumber\\
 &-& 
   4\sto{\tilde{U}}{p}{q}{q}
     \left\{
     \vphantom{\frac{G_{ep}^2}{2q^2}}
      \frac{mG_{ee}}{q^3}
      I\left[\frac{{\mathbf pq}}{q} 
       \right]
    \right. \nonumber\\
 & &\qquad~~~~~~~~
     + \frac{G_{ep}^2}{2q^2}\frac{m^2s/S}{q^2+\lambda^2}
        \left[\frac1{q^2 - 2mSq} - \frac1{q^2 + 2mSq}
        \right]
      I\left[\frac{{\mathbf pq}}{q}\right]
 \nonumber \\
 & &\qquad~~~~~~~~
     + \frac{G_{ep}^2}{2q^2}\frac{m^2s/S}{q^2+\lambda^2}
        \left[\frac{1}{2{\mathbf pq} + q^2 - 2mSq}
        \right]
        \left(I\left[\frac{{\mathbf pq}}{q}
               \right]
            - I\left[mS-\frac{q}{2}
               \right]
        \right)
 \nonumber \\
 & &\qquad~~~~~~~~\left.
     - \frac{G_{ep}^2}{2q^2}\frac{m^2s/S}{q^2+\lambda^2}
        \left[\frac{1}{2{\mathbf pq} + q^2 + 2mSq}
        \right]
        \left(I\left[\frac{{\mathbf pq}}{q}
               \right]
            + I\left[mS+\frac{q}{2}
               \right]
        \right)
     \right\}
 \nonumber \\
 &+& 
   4\sto{\tilde{U}}{-q}{q}{q}
     \left[ \frac{mG_{ee}}{q^3}
          + \frac{G_{ep}^2}{2q^2}\frac{m^2s/S}{q^2+\lambda^2}
             \left(\frac1{2{\mathbf pq} + q^2 - 2mSq}
                 - \frac1{2{\mathbf pq} + q^2 + 2mSq}
             \right)
     \right] I[q]
 \ .\label{eq-50}
\end{eqnarray}

  Now, it is necessary to make a remark. 
We will assume that $p\approx K_F$.
In other words, we are interested in the interaction 
of electrons in the vicinity of the Fermi surface.
Then, at $q \to 0$,
the fractions of the type
$\frac1{2{\mathbf pq} + q^2 \pm 2mSq}$
actually contain a factor  $K_F^{-1}$, 
whereas the fractions of the type
$\frac1{q^2 \pm 2mSq}$ contain the factor $(mS)^{-1}$.
So, the fractions of the first type turn out to be small
as compared to the fractions of the second type
with the order of smallness $\sim mS/K_F \ll 1$. 

  These arguments provide a basis to exclude from Eq.~(\ref{eq-50})
terms containing the fractions of the type 
$\frac1{2{\mathbf pq} + q^2 \pm 2mSq}$ 
as factors.
This will simplify Eq.~(\ref{eq-50}), 
and, after proper transformation, it can be reduced to the form 
\begin{eqnarray}
 \sto{\tilde{U}}{p}{q}{q}
 &=& \sto{U}{p}{q}{q}
   + 4\frac{mG_{ee}}{q^3}I[q]\sto{\tilde{U}}{-q}{q}{q}
 \nonumber\\
 & & \qquad~~~~~~
   + 4\sto{\tilde{U}}{p}{q}{q}
    I'\left[\frac{{\mathbf pq}}{q}
      \right]
     \left[
       \frac{mG_{ee}}{q^2}
     + \frac{G_{ep}^2}{2q^2}\frac{m^2s/S}{q^2+\lambda^2}
        \left(\frac1{q - 2mS} - \frac1{q + 2mS}
        \right)
     \right]
 \ ,\label{eq-51}
\end{eqnarray}
where, within the limits of our rough approximations, 
it is possible to assume 
\begin{equation}
  \left.
  \left(I\left[\frac{({\mathbf p+q}){\mathbf q}}{q}\right]
      - I\left[\frac{{\mathbf pq}}{q}\right]
  \right)
  \right/q \approx
  I'\left[\frac{{\mathbf pq}}{q}\right]
  \ ;\quad
  I'[x] \equiv \frac{d}{dx}I[x]
 \ .\nonumber
\end{equation}
Leaving at $q \to 0$ merely the main terms 
and in the view of~(\ref{eq-49})
it is possible to derive from Eq.~(\ref{eq-51})
the following one:
\begin{equation}
 4\frac{mG_{ee}}{q^2}
  I'\left[\frac{{\mathbf pq}}{q}
    \right]
   \sto{\tilde{U}}{p}{q}{q}
 =- \frac{G_{ee}}{2q^2}
  - 2\left(\frac{zm}{3M}\right)^2
      \frac{G_{ee}}{q^2}
       \frac{K_F^2}{q^2-\chi_1}
        \frac{K_F^2}{q^2-\chi}
 \label{eq-52}
\end{equation}

  In an analogous way for $\sto{\tilde{U}}{p}{k}{q}$
at $p, k \approx K_F$
accounting for the remarks to derivation of Eq.(\ref{eq-51})
we obtain
\begin{equation}
 \sto{\tilde{U}}{p}{k}{q}
 =  \frac{G_{ee}}{q^2}
 + 4\frac{mG_{ee}}{q^2}
    I'\left[\frac{{\mathbf pq}}{q}
      \right]
    \sto{\tilde{U}}{p}{q}{q}
 + 4\frac{mG_{ee}}{q^2}
    I'\left[\frac{{\mathbf kq}}{q}
      \right]
    \sto{\tilde{U}}{k}{-q}{-q}
 \label{eq-53}
\end{equation}
and finally
\begin{equation}
 \sto{\tilde{U}}{p}{k}{q}
 =- 4 \left(\frac{zm}{3M}\right)^2
      \frac{G_{ee}}{q^2}
       \frac{K_F^2}{q^2-\chi_1}
        \frac{K_F^2}{q^2-\chi}
 \ .\label{eq-54}
\end{equation}

In a coordinate representation el-el interaction 
$V(\mathbf r)$ may be estimated as follows
($Q^2=\chi$, $Q_1^2=\chi_1$): 
\begin{equation}
  V({\mathbf r}) 
  \approx \left(\frac{zm}{M}\right)^2
      \frac{G_{ee}K_F^4}{\left(\chi_1-\chi\right)r}
       \left[\chi_1^{-1}\sin^2\left(\frac{Q_1r}2\right)
           - \chi^{-1}\sin^2\left(\frac{Qr}2\right)
       \right]
 \ .\label{eq-55}
\end{equation}
    
  According to~(\ref{eq-54}) the specific features 
of the el-el interaction associated with 
the virtual phonon exchange 
result in the absence of screening of the long-wave part 
of this interaction component.
We would like to note here that 
in the case of Coulomb el-el interaction the same calculation procedure
(Appendix 4) gives the result, 
which is consistent with 
the consideration of electron plasma in Thomas--Fermi approximation. 

  The potential of the type~(\ref{eq-55}) causes instability 
with respect to formation of pairs, 
which may lead to a superconductive phase transition.
Without taking in consideration short-wave corrections
the pair binding energy $E_b$ can be estimated 
as follows:
\begin{equation}
  E_b \sim \left(\frac{zm}{M}\right)^2\frac{G_{ee}K_F^4}{Q^3}
 \ .\label{eq-56}
\end{equation} 

  Several remarks should be made. 
Firstly, it is obvious that $E_b\sim M^{-1/2}$,
that is in agreement with the observations 
of the isotopic effect in a series of metals~\cite{bib-19}.
Secondly, the term $E_b$ in Eq.~(\ref{eq-56})
has been explicitly overrated 
if we have the intention to estimate $T_s$
from $E_b$ according to standard procedure.
In fact, Eq.~(\ref{eq-56}) determines the energy gap
averagely over the Fermi surface while 
the transition temperature is determined 
by its minimum value over it.
This means that enhancing symmetry should lead 
to an increase in $T_s$,
which is consistent with the observation of superconductivity 
in amorphous alloys~\cite{bib-20}.
Thirdly, and it is the main point,
oscillations of the potential $V({\mathbf r})$
lead to the decrease of interaction between pairs.
This is due to the fact
that calculating the corresponding matrix elements
involves some averaging procedure of $V({\mathbf r})$
over the regions roughly equal in size to the size of a pair.
This question needs detailed consideration and 
is postponed for further publication. 

  Of particular interest is the singular behavior 
of the binding energy $E_b $ at $Q \to 0$, 
following from Eq.~(\ref{eq-56}).
In real systems this singularity should not take place
in the behavior of $T_s $ 
even though in virtue of the symmetry reasons stated above. 
Nevertheless maximum values of $T_s$ should be observed 
if the following condition were fulfilled:
\begin{equation}
 \tilde{S}^2 \approx \frac{zm}{3M} V_F^2
 \ .\label{eq-57}
\end{equation} 

  This condition may be fulfilled with sufficient accuracy in systems 
with widely modified physical parameters, for instance,
in solid solutions, where we believe HTSC is to be observed
when the condition~(\ref{eq-57}) is fulfilled.
From this viewpoint the researches into HTSC 
in stoichemetric compounds of simple composition~\cite{bib-14,bib-15,bib-16}
become of current interest because reliable zone 
calculations can be carried out just for these materials.
In this case analyzing experimental data 
it should be kept in mind that the parameters $M$ and $z$ in Eq.~(\ref{eq-57})
become equal to the sum of masses of ions and number of conductivity electrons
respectively per cell. 
\newpage
\begin{appendix}
\vspace*{12mm}
\appendix
\hspace*{2mm}{\Large \bf Appendix}
\vspace{-3mm}
\section{Invariance of the commutation relations for 
 creation and annihilation operators with respect to
 transformations~(9)-(10).}
\renewcommand{\theequation}{\thesection.\arabic{equation}}
 \label{App-1}
  \setcounter{equation}{0}

  For the commutator $\left[\scc{\mu}{p},\scn{\nu}{k}\right]$
we have:
\begin{eqnarray}
 \left[\scc{\mu}{p};\scn{\nu}{k}\right] 
 &=& \left[\ccc{\mu}{p};\ccn{\nu}{k}\right]
 \nonumber \\
 &-& \int \fic{k}{q}
      \left[\ccc{\mu}{p};\ccn{\nu}{k-q}\bbn{q}\right]
       d{\mathbf q}
   + \int \fin{k+q}{q}
      \left[\ccc{\mu}{p};\ccn{\nu}{k+q}\bbc{q}\right]
       d{\mathbf q}
 \nonumber \\
 &-& \frac12 \int \left|\fin{k}{q}\right|^2
      \left[\ccc{\mu}{p};\ccn{\nu}{k}\right]
       d{\mathbf q}
 \nonumber \\
 &+& \frac12 \sum_{\xi}\int\!\!\!\int
      \left( \fic{m}{-q}\fin{k-q}{-q}
           - \fic{k}{q}\fin{m+q}{q}
      \right)
       \left[\ccc{\mu}{p};\ccc{\xi}{m}\ccn{\xi}{m+q}\ccn{\nu}{k-q}\right]
        d{\mathbf q} d{\mathbf m}
 \nonumber \\
 &-& \frac12 \int\!\!\!\int
      \left( \fic{k}{m}\fin{k-m+q}{q}
           + \fic{k+q}{m}\fin{k+q}{q}
      \right)
       \left[\ccc{\mu}{p};\ccn{\nu}{k-m+q}\bbc{q}\bbn{m}\right]
        d{\mathbf q} d{\mathbf m}
 \nonumber \\
 &+& \frac14 \int\!\!\!\int
      \left( \fic{k-m}{q}\fic{k}{m}
           + \fic{k-q}{m}\fic{k}{q}
      \right)
       \left[\ccc{\mu}{p};\ccn{\nu}{k-m-q}\bbn{q}\bbn{m}\right]
        d{\mathbf q} d{\mathbf m}
 \nonumber \\
 &+& \frac14 \int\!\!\!\int
      \left( \fin{k+q}{q}\fin{k+m+q}{m}
           + \fin{k+m}{m}\fin{k+m+q}{q}
      \right)
       \left[\ccc{\mu}{p};\ccn{\nu}{k+m+q}\bbc{q}\bbc{m}\right]
        d{\mathbf q} d{\mathbf m}
 \nonumber \\
 &-& \int \fin{p}{q}
      \left[\ccc{\mu}{p-q}\bbc{q};\ccn{\nu}{k}\right]
       d{\mathbf q}
   + \int \fic{p+q}{q}
      \left[\ccc{\mu}{p+q}\bbn{q};\ccn{\nu}{k}\right]
       d{\mathbf q}
 \nonumber \\
 &-& \frac12 \int \left|\fin{p}{q}\right|^2
      \left[\ccc{\mu}{p};\ccn{\nu}{k}\right]
       d{\mathbf q}
 \nonumber \\
 &+& \frac12 \sum_{\xi}\int\!\!\!\int
      \left( \fic{p-q}{-q}\fin{m}{-q}
           - \fic{m+q}{q}\fin{p}{q}
      \right)
       \left[\ccc{\mu}{p-q}\ccc{\xi}{m+q}\ccn{\xi}{m};\ccn{\nu}{k}\right]
        d{\mathbf q} d{\mathbf m}
 \nonumber \\
 &-& \frac12 \int\!\!\!\int
      \left( \fic{p-m+q}{q}\fin{p}{m}
           + \fic{p+q}{q}\fin{p+q}{m}
      \right)
       \left[\ccc{\mu}{p-m+q}\bbc{m}\bbn{q};\ccn{\nu}{k}\right]
        d{\mathbf q} d{\mathbf m}
 \nonumber \\
 &+& \frac14 \int\!\!\!\int
      \left( \fin{p-m}{q}\fin{p}{m}
           + \fin{p-q}{m}\fin{p}{q}
      \right)
       \left[\ccc{\mu}{p-m-q}\bbc{q}\bbc{m};\ccn{\nu}{k}\right]
        d{\mathbf q} d{\mathbf m}
 \nonumber \\
 &+& \frac14 \int\!\!\!\int
      \left( \fic{p+q}{q}\fic{p+m+q}{m}
           + \fic{p+m}{m}\fic{p+m+q}{q}
      \right)
       \left[\ccc{\mu}{p+m+q}\bbn{q}\bbn{m};\ccn{\nu}{k}\right]
        d{\mathbf q} d{\mathbf m}
 \nonumber \\
 &+& \int\!\!\!\int
      \left\{
        \fic{p}{q}\fin{k}{m}
         \left[\ccc{\mu}{p-q}\bbc{q};\ccn{\nu}{k-m}\bbn{m}\right]
      - \fin{p}{q}\fin{k+m}{m}
         \left[\ccc{\mu}{p-q}\bbc{q};\ccn{\nu}{k+m}\bbc{m}\right]
      \right\}
       d{\mathbf q} d{\mathbf m}
 \nonumber \\
 &=& \left[\ccc{\mu}{p};\ccn{\nu}{k}\right]
     \left( 1 - \int \left|\fin{p}{q}\right|^2 d{\mathbf q}
            +   \int \left|\fin{p}{q}\right|^2 d{\mathbf q}
     \right)
 \nonumber \\
 &+& \delta_\nu^\mu
      \left[ \fin{p}{p-k} \bbc{p-k} - \fic{k}{k-p} \bbn{k-p}
           - \fin{p}{p-k} \bbc{p-k} + \fic{k}{k-p} \bbn{k-p}
      \right]
 \nonumber \\
 &+& \frac12 \delta_\nu^\mu \int
      \left( \fic{q}{p-k}\fin{p}{p-k} - \fic{k}{k-p}\fin{k+q-p}{k-p}
      \right)
       \sum_{\xi} \ccc{\xi}{q}\ccn{\xi}{k+q-p}
        d{\mathbf q}
 \nonumber \\
 &+& \frac12 \delta_\nu^\mu \int
      \left( \fic{k}{k-p}\fin{q}{k-p}
           - \fic{p-k+q}{p-k}\fin{p}{p-k}
      \right)
       \sum_{\xi}\ccc{\xi}{p-k+q}\ccn{\xi}{q}
        d{\mathbf q}
 \nonumber \\
 &-& \frac12 \int
      \left( \fic{p-q}{-q}\fin{k-q}{-q} - \fic{k}{q}\fin{p}{q}
      \right)
       \ccc{\mu}{p-q}\ccn{\nu}{k-q}
        d{\mathbf q}
 \nonumber \\
 &-& \frac12 \int
      \left( \fic{p-q}{-q}\fin{k-q}{-q}
           - \fic{k}{q}\fin{p}{q}
      \right)
       \ccc{\mu}{p-q}\ccn{\nu}{k-q}
        d{\mathbf q}
 \nonumber \\
 &+& \int
      \left[ \fic{p+q}{q}\fin{k+q}{q}\ccc{\mu}{p+q}\ccn{\nu}{k+q}
           - \fin{p}{q}\fic{k}{q}\ccc{\mu}{p-q}\ccn{\nu}{k-q}
      \right]
       d{\mathbf q}
 \nonumber \\
 &-& \frac12 \delta_\nu^\mu \int
      \left( \fic{k}{k-p+q}\fin{p}{q} + \fic{k+q}{k-p+q}\fin{k+q}{q}
      \right)
       \bbc{q}\bbn{k-p+q}
        d{\mathbf q}
 \nonumber \\
 &-& \frac12 \delta_\nu^\mu \int
      \left( \fic{k}{q}\fin{p}{p-k+q}
           + \fic{p+q}{q}\fin{p+q}{p-k+q}
      \right)
       \bbc{p-k+q}\bbn{q}
        d{\mathbf q}
 \nonumber \\
 &+& \delta_\nu^\mu \int
      \left[ \fin{p}{q}\fic{k}{k-p+q}\bbc{q}\bbn{k-p+q}
           + \fic{p+q}{q}\fin{p+q}{p-k+q}\bbc{p-k+q}\bbn{q}
      \right]
       d{\mathbf q}
 \nonumber \\
 &+& \frac14 \delta_\nu^\mu \int
      \left( \fic{p+q}{q}\fic{k}{k-p-q} + \fic{k-q}{k-p-q}\fic{k}{q}
      \right)
       \bbn{q}\bbn{k-p-q}
        d{\mathbf q}
 \nonumber \\
 &+& \frac14 \delta_\nu^\mu \int
      \left( \fic{p+q}{q}\fic{k}{k-p-q}
           + \fic{k-q}{k-p-q}\fic{k}{q}
      \right)
       \bbn{q}\bbn{k-p-q}
        d{\mathbf q}
 \nonumber \\
 &-& \delta_\nu^\mu \int
      \left[ \fic{p+q}{q}\fic{k}{k-p-q}\bbn{q}\bbn{k-p-q}
           + \fin{p}{q}\fin{p-q}{p-k-q}\bbc{q}\bbc{p-k-q} 
      \right]
       d{\mathbf q}
 \nonumber \\
 &+& \frac14 \delta_\nu^\mu \int
      \left( \fin{k+q}{q}\fin{p}{p-k-q} + \fin{p-q}{p-k-q}\fin{p}{q}
      \right)
       \bbc{q}\bbc{p-k-q}
        d{\mathbf q}
 \nonumber \\
 &+& \frac14 \delta_\nu^\mu \int
      \left( \fin{k+q}{q}\fin{p}{p-k-q}
           + \fin{p-q}{p-k-q}\fin{p}{q}
      \right)
       \bbc{q}\bbc{p-k-q}
        d{\mathbf q}
  \nonumber
\end{eqnarray}

After proper changing of integration variable $q$ 
in some integrals it becomes clear that
\begin{equation}
 \left[\scc{\mu}{p};\scn{\nu}{k}\right] 
 = \left[\ccc{\mu}{p};\ccn{\nu}{k}\right]
 \nonumber 
\end{equation} 
with an accuracy up to terms $\sim |\varphi|^3$.
The invariance for the remaining commutation relations
$\left[\scc{\mu}{p};\scc{\nu}{k}\right]$,
$\left[\scc{\mu}{p};\sbc{k}\right]$,
$\left[\scc{\mu}{p};\sbn{k}\right]$,
$\left[\sbc{p};\sbn{k}\right]$, and
$\left[\sbc{p};\sbc{k}\right]$
is proved in close analogy to what has been given above.

\section{Invariance of the commutation relations for 
 creation and annihilation operators with respect to
 transformations~(33).}
 \label{App-2}
  \setcounter{equation}{0}

  We have:
\begin{eqnarray}
 \left[\ccc{\mu}{p}\ ;\ \ccn{\nu}{k}\right] 
 &=& \left[\tcc{\mu}{p}
           -\sum_{\xi} \int\!\!\!\int
            \sto{\theta_{\xi}^{\mu}}{p}{m}{q}
             \tcc{\mu}{p+q}\tcc{\xi}{m-q}\tcn{\xi}{m}
              d{\mathbf m} d{\mathbf q}
     \right.\ ;\nonumber \\ &&\left.~~
          \tcn{\nu}{k}
          -\sum_{\xi} \int\!\!\!\int
           \sto{\theta_{\xi}^{\nu}}{k}{n}{s}
            \tcc{\xi}{n}\tcc{\xi}{n-s}\tcn{\nu}{k+s}
             d{\mathbf n} d{\mathbf s}
     \right]
 \nonumber \\
 &=& \left[\tcc{\mu}{p}\ ;\ \tcn{\nu}{k}\right]
  - \sum_{\xi} \int\!\!\!\int
     \sto{\theta_{\xi}^{\nu}}{k}{n}{s}
      \left[\tcc{\mu}{p}\ ;\
            \tcc{\xi}{n}\tcc{\xi}{n-s}\tcn{\nu}{k+s}
            d{\mathbf n} d{\mathbf s}
      \right]  \nonumber \\ && \qquad \qquad \quad ~
  - \sum_{\xi} \int\!\!\!\int
     \sto{\theta_{\xi}^{\mu}}{p}{m}{q}
      \left[\tcc{\mu}{p+q}\tcc{\xi}{m-q}\tcn{\xi}{m}\ ;\
            \tcn{\nu}{k}
            d{\mathbf m} d{\mathbf q}
      \right]
  + O\left(\theta^2\right)\ .
 \nonumber 
\end{eqnarray}
Opening the commutators
$\left[\tcc{\mu}{p}\ ;\ \tcc{\xi}{n}\tcc{\xi}{n-s}\tcn{\nu}{k+s}\right]$
and 
$\left[\tcc{\mu}{p+q}\tcc{\xi}{m-q}\tcn{\xi}{m}\ ;\ \tcn{\nu}{k}\right]$
we derive 
\begin{eqnarray}
 \left[\ccc{\mu}{p}\ ;\ \ccn{\nu}{k}\right] 
 &=& \left[\tcc{\mu}{p}\ ;\ \tcn{\nu}{k}\right]
  \nonumber \\ 
 &-& \delta_{\nu}^{\mu} \sum_{\xi} \int
      \left[\sto{\theta_{\xi}^{\nu}}{k}{q}{p-k}
          + \sto{\theta_{\xi}^{\mu}}{p}{k-p+q}{k-p}
      \right]
       \tcc{\xi}{q}\tcn{\xi}{k-p+q}
        d{\mathbf q}
  \nonumber \\ 
 &+& \int
      \left[\sto{\theta_{\mu}^{\nu}}{k}{q}{q-p}
          + \sto{\theta_{\nu}^{\mu}}{p}{k-p+q}{q-p}
      \right]
       \tcc{\mu}{q}\tcn{\nu}{k-p+q}
        d{\mathbf q}
   + O\left(\theta^2\right)\ .
 \nonumber 
\end{eqnarray}
Substituting the term $\theta$ from Eq.(\ref{eq-36})
in view of Eqs.(\ref{eq-37}) and (\ref{eq-40}) 
we definitely have
\begin{equation}
 \left[\ccc{\mu}{p}\ ;\ \ccn{\nu}{k}\right] 
 = \left[\tcc{\mu}{p}\ ;\ \tcn{\nu}{k}\right]
\end{equation}
with an accuracy up to terms $\sim \theta^2$.

 The invariance for the commutation  relations
$\left[\tcc{\mu}{p}\ ;\ \tcc{\nu}{k}\right]$ and
$\left[\tcn{\mu}{p}\ ;\ \tcn{\nu}{k}\right]$
is proved in close analogy to what has been given above.

\section{Testing diagonalization of the Hamiltonian~(36)
         by the transformation~(33).}
 \label{App-3}
  \setcounter{equation}{0}
  Expressing Eq.~(\ref{eq-38}) via operators $\tcc{}{}$
and $\tilde{C}$
it is possible to obtain with an accuracy up to terms $\sim \theta^2$ 
\begin{eqnarray}
&& \int \epp{p}
    \left[\tcc{\mu}{p}
        - \int \sum_{\nu} \sto{\theta_{\nu}^{\mu}}{p}{k}{q}
           \tcc{\mu}{p+q}\tcc{\nu}{k-q}\tcn{\nu}{k}
            d{\mathbf k} d{\mathbf q}
    \right]
    \left[\tcn{\mu}{p}
        - \int \sum_{\nu} \sto{\theta_{\nu}^{*\mu}}{p}{k}{q}
           \tcc{\nu}{k}\tcn{\nu}{k-q}\tcn{\mu}{p+q}
            d{\mathbf k} d{\mathbf q}
    \right]
     d{\mathbf p}
 \nonumber \\ &=&
   \int \epp{p} \tcc{\mu}{p} \tcn{\mu}{p}
  - \int\!\!\!\int\!\!\!\int \sum_{\nu} \epp{p}
     \sto{\theta_{\nu}^{*\mu}}{p}{k}{q}
      \tcc{\mu}{p}\tcc{\nu}{k}
       \tcn{\nu}{k-q}\tcn{\mu}{p+q}
        d{\mathbf k} d{\mathbf q} d{\mathbf p}
 \nonumber \\ &&\qquad\qquad\quad~~
  - \int\!\!\!\int\!\!\!\int \sum_{\nu} \epp{p} 
     \sto{\theta_{\nu}^{\mu}}{p}{k}{q}
      \tcc{\mu}{p+q}\tcc{\nu}{k-q}
       \tcn{\nu}{k}\tcn{\mu}{p}
        d{\mathbf k} d{\mathbf q} d{\mathbf p}
\ +\ O(\theta^2)
\ .\label{eq-C1}
\end{eqnarray}
   Exchanging variables $p'=p+q$, $k'=k-q$, and $q'=-q$ we obtain 
$$
 \int\!\!\!\int\!\!\!\int \epp{p}
  \sto{\theta_{\nu}^{*\mu}}{p}{k}{q}
   \tcc{\mu}{p}\tcc{\nu}{k}
    \tcn{\nu}{k-q}\tcn{\mu}{p+q}
    d{\mathbf k} d{\mathbf q} d{\mathbf p}
= \int\!\!\!\int\!\!\!\int \epp{p}  
   \sto{\theta_{\nu}^{*\mu}}{p+q}{k-q}{-q}
    \tcc{\mu}{p+q}\tcc{\nu}{k-q}
     \tcn{\nu}{k}\tcn{\mu}{p}
      d{\mathbf k} d{\mathbf q} d{\mathbf p}
$$
and then $H_{\theta}$,
the sum of linear in $\theta$ terms in Eq.(\ref{eq-C1}),
is
$$
H_{\theta} =
 - \int\!\!\!\int\!\!\!\int \sum_{\nu} 
    \left[\epp{p+q}\sto{\theta_{\nu}^{*\mu}}{p+q}{k-q}{-q}
        + \epp{p}\sto{\theta_{\nu}^{\mu}}{p}{k}{q}
    \right]
     \tcc{\mu}{p+q}\tcc{\nu}{k-q}
      \tcn{\nu}{k}\tcn{\mu}{p}
       d{\mathbf k} d{\mathbf q} d{\mathbf p}
 \ .
$$

 The second substitution of variables $k'=p$, $p'=k$, and $q'=-q$
gives the following expression
\begin{eqnarray}
H_{\theta} &=&
 - \frac12
    \int\!\!\!\int\!\!\!\int \sum_{\nu} 
     \left[ \epp{p+q}\sto{\theta_{\nu}^{*\mu}}{p+q}{k-q}{-q}
          + \epp{p}\sto{\theta_{\nu}^{\mu}}{p}{k}{q}
          + \epp{k-q}\sto{\theta_{\nu}^{*\mu}}{k-q}{p+q}{q}
          + \epp{k}\sto{\theta_{\nu}^{\mu}}{k}{p}{-q}
     \right]\times \nonumber\\
&&\qquad\qquad\qquad\qquad\qquad\qquad\qquad\qquad\qquad\qquad\qquad\quad
  \times\
   \tcc{\mu}{p+q}\tcc{\nu}{k-q}
    \tcn{\nu}{k}\tcn{\mu}{p}
     d{\mathbf k} d{\mathbf q} d{\mathbf p}
 \ . \nonumber
\end{eqnarray}
Substituting $\theta$ by the term~(\ref{eq-36})
in view of Eqs.(\ref{eq-37}) and (\ref{eq-40})
leads to the following relation
$$ H_{\theta} =
 - \int\!\!\!\int\!\!\!\int \sum_{\nu} 
    \left\{ \sto{\tilde{U}}{p}{k}{q}
         + \frac12\delta_{\nu}^{\mu}
            \left[2\sto{\tilde{V}}{p}{k}{q}
                 - \sto{\tilde{U}}{p}{k}{q}
                 - \sto{\tilde{U}}{p}{k}{k-p-q}
            \right]
     \right\}
      \tcc{\mu}{p+q}\tcc{\nu}{k-q}
       \tcn{\nu}{k}\tcn{\mu}{p}
        d{\mathbf k} d{\mathbf q} d{\mathbf p}\ .
$$

  On the other hand, 
by separate considering cases $\mu=\nu$ and $\mu\neq\nu$
it is possible to obtain the following expression
for the term $\tilde{H}_{ee}$ in Eq.(\ref{eq-39}): 
$$
 \tilde{H}_{ee} =
 \int\!\!\!\int\!\!\!\int
  \sum_{\mu}\sum_{\nu} 
   \left\{ \sto{\tilde{U}}{p}{k}{q}
         + \frac12\delta_{\nu}^{\mu}
            \left[2\sto{\tilde{V}}{p}{k}{q}
                 - \sto{\tilde{U}}{p}{k}{q}
                 - \sto{\tilde{U}}{p}{k}{k-p-q}
            \right]
   \right\}
    \tcc{\mu}{p+q}\tcc{\nu}{k-q}
     \tcn{\nu}{k}\tcn{\mu}{p}
      d{\mathbf k} d{\mathbf q} d{\mathbf p}\ .
$$

 Thus, it is the transformation~(\ref{eq-35})
that diagonalizes the term $H_{\textrm tot}$~(\ref{eq-38})
with an accuracy up to terms $\sim \tilde{U}^2, \tilde{V}^2$.

\section{Plasma correlations in the case of Coulombs's
         el-el interaction}
 \label{App-4}
  \setcounter{equation}{0}

  Let us assume in this case that the effective interaction 
$\sto{\tilde{U}}{p}{k}{q}$ is described by quite a regular function 
of all three parameters {\bf p}, {\bf k}, and {\bf q}.
Then the equation determining the term $\sto{\tilde{U}}{p}{k}{q}$
has the following form: 
\begin{eqnarray}
 \sto{\tilde{U}}{p}{k}{q}
 &=& \frac{G_{ee}}{q^2}
 \nonumber\\
 &+& \frac{mG_{ee}}{q^3}
      \left\{
        \left[2\sto{\tilde{U}}{q}{p}{p}
            - 4\sto{\tilde{U}}{p}{q}{q}
        \right]
         I\left[\frac{{\mathbf pq}}{q}\right]
      + \left[4\sto{\tilde{U}}{p+q}{-q}{-q}
            - 2\sto{\tilde{U}}{-q}{p+q}{p+q}
        \right]
         I\left[\frac{{\mathbf pq}}{q}+q\right]
      \right.
 \nonumber\\
 & &  \left.\qquad~
      + \left[4\sto{\tilde{U}}{k}{-q}{-q}
            - 2\sto{\tilde{U}}{-q}{k}{k}
        \right]
         I\left[\frac{{\mathbf kq}}{q}\right]
      + \left[2\sto{\tilde{U}}{q}{k-q}{k-q}
           - 4\sto{\tilde{U}}{k-q}{q}{q}
        \right]
        I\left[\frac{{\mathbf kq}}{q}-q\right]
      \right\}
 \ .\label{eq-D1}
\end{eqnarray}

  By direct substitution into Eq.(\ref{eq-D1})
it is possible to show that 
$\sto{\tilde{U}}{p+q}{-q}{-q}=\sto{\tilde{U}}{p}{q}{q}$ and
$\sto{\tilde{U}}{k}{-q}{-q}=\sto{\tilde{U}}{k-q}{q}{q}$.
Therefore, Eq.(\ref{eq-D1}) is reduced to the form
\begin{eqnarray}
 \sto{\tilde{U}}{p}{k}{q}
 = \frac{G_{ee}}{q^2}
   + \frac{mG_{ee}}{q^3}
      \left\{
        4\left(I\left[\frac{{\mathbf pq}}{q}+q\right]
             - I\left[\frac{{\mathbf pq}}{q}\right]
         \right)\sto{\tilde{U}}{p}{q}{q}
      + 4\left(I\left[\frac{{\mathbf kq}}{q}\right]
             - I\left[\frac{{\mathbf kq}}{q}-q\right]
         \right)\sto{\tilde{U}}{k}{-q}{-q}
      \right.
 \nonumber\\
\left.+\ 2I\left[\frac{{\mathbf pq}}{q}\right]
         \sto{\tilde{U}}{q}{p}{p}
      - 2I\left[\frac{{\mathbf pq}}{q}+q\right]
         \sto{\tilde{U}}{-q}{p+q}{p+q}
      - 2I\left[\frac{{\mathbf kq}}{q}\right]
         \sto{\tilde{U}}{-q}{k}{k}
      + 2I\left[\frac{{\mathbf kq}}{q}-q\right]
         \sto{\tilde{U}}{q}{k-q}{k-q}
      \right\}
 \ .\label{eq-D2}
\end{eqnarray}

  Let us consider the calculation of the term 
$\sto{\tilde{U}}{-q}{q}{q}$:
\begin{equation}
 \sto{\tilde{U}}{-q}{q}{q}
 = \frac{G_{ee}}{q^2}
 + 4\frac{mG_{ee}}{q^3}
    I[q]\sto{\tilde{U}}{-q}{q}{q}
 \ ;\qquad
 \sto{\tilde{U}}{-q}{q}{q}
 = \frac{qG_{ee}}{q^3 - 4mG_{ee}I[q]}
\ .\label{eq-D3}
\end{equation}
Note that at $q \to 0$ we have
\begin{equation}
 \sto{\tilde{U}}{-q}{q}{q}
 = \frac{G_{ee}}{q^2 + \lambda^2}
\ ,\label{eq-D4}
\end{equation}
whereas at $q\to K_F$ this term is logarithmically small:
$\sto{\tilde{U}}{-q}{q}{q} \sim \ln^{-1}\left|q-K_F\right|$.
 
  Using Eqs.(\ref{eq-D2}) and (\ref{eq-D3}) it is possible to obtain
\begin{eqnarray}
 \left[1
     + 4\frac{mG_{ee}}{q^3}
        \left(I\left[\frac{{\mathbf pq}}{q}\right]
            - I\left[\frac{{\mathbf pq}}{q}+q\right]
        \right)
 \right]\sto{\tilde{U}}{p}{q}{q}
 &=& \frac{G_{ee}}{2q^2}
 + \frac12 \sto{\tilde{U}}{-q}{q}{q}
 + 2\frac{mG_{ee}}{q^3}\times\nonumber\\
 &&
    \left(I\left[\frac{{\mathbf pq}}{q}\right]
           \sto{\tilde{U}}{q}{p}{p}
        - I\left[\frac{{\mathbf pq}}{q}+q\right]
           \sto{\tilde{U}}{-q}{p+q}{p+q}
    \right)
 \ ;\label{eq-D5}\\
 \left[1
     + 4\frac{mG_{ee}}{q^3}
        \left(I\left[\frac{{\mathbf kq}}{q}-q\right]
            - I\left[\frac{{\mathbf kq}}{q}\right]
        \right)
 \right]\sto{\tilde{U}}{k}{-q}{-q}
 &=& \frac{G_{ee}}{2q^2}
 + \frac12 \sto{\tilde{U}}{-q}{q}{q}
 - 2\frac{mG_{ee}}{q^3}\times\nonumber\\
 &&
    \left(I\left[\frac{{\mathbf kq}}{q}\right]
           \sto{\tilde{U}}{-q}{k}{k}
        - I\left[\frac{{\mathbf kq}}{q}-q\right]
           \sto{\tilde{U}}{q}{k-q}{k-q}
    \right)
 \ .\label{eq-D6}
\end{eqnarray}
Comparing Eqs.(\ref{eq-D2}) and (\ref{eq-D5})-(\ref{eq-D6})
makes it evident that
\begin{equation}
 \sto{\tilde{U}}{p}{k}{q} =
 \sto{\tilde{U}}{p}{q}{q} +
 \sto{\tilde{U}}{k}{-q}{-q} -
 \sto{\tilde{U}}{-q}{q}{q}
 \ .\label{eq-D7}
\end{equation}

  So, the determination of the interaction potential reduces 
to the calculation of the terms
$\sto{\tilde{U}}{p}{q}{q}$ and $\sto{\tilde{U}}{k}{-q}{-q}$.
Further we will limit our investigation to the case 
of $p,\ k \approx K_F$. 

  Considering Eq.(\ref{eq-D5}) ensures us 
that the terms $\sto{\tilde{U}}{q}{p}{p}$ and 
$\sto{\tilde{U}}{-q}{p+q}{p+q}$ are logarithmically small
as well as the term $\sto{\tilde{U}}{-p}{p}{p}$ is.
This permits us to exclude the terms 
$\sto{\tilde{U}}{q}{p}{p}$ and 
$\sto{\tilde{U}}{-q}{p+q}{p+q}$ from Eq.(\ref{eq-D5}).
Then we have: 
\begin{equation}
 \left[1 - 4\frac{mG_{ee}}{q^2}
            I'\left[\frac{{\mathbf pq}}{q}\right]
 \right]\sto{\tilde{U}}{p}{q}{q} =
 \frac{G_{ee}}{2q^2}
 + \frac12 \sto{\tilde{U}}{-q}{q}{q}
 \ .\label{eq-D8}
\end{equation}

  For the term $I'$ we use the expression
$$ I'[x] =
 - \pi K_F^2\left(\frac{1}{K_F-x}
                + \frac{1}{K_F+x}
           \right)
 - 2 \pi K_F =
 - 2 \pi K_F \frac{2K_F^2-x^2}{K_F^2-x^2}\ .
$$
Singling out the main terms in Eq.(\ref{eq-D8})
at $q \to 0$, we come to the following equation
for the term $\sto{\tilde{U}}{p}{q}{q}$
\begin{equation}
 \left[1 + \frac{\lambda^2}{2q^2}
            \frac{2q^2K_F^2-({\mathbf pq})^2}{q^2K_F^2-({\mathbf pq})^2}
 \right]\sto{\tilde{U}}{p}{q}{q} =
 \frac{G_{ee}}{2q^2}
 \ ,\label{eq-D9}
\end{equation}
and consequently 
\begin{equation}
\sto{\tilde{U}}{p}{q}{q} =
 G_{ee}\frac{q^2K_F^2-({\mathbf pq})^2}
            {2q^2K_F^2\left(q^2+\lambda^2\right)
             -({\mathbf pq})^2\left(2q^2+\lambda^2\right)}
 \ .\label{eq-D10}
\end{equation}
Analogously we get
\begin{equation}
\sto{\tilde{U}}{k}{-q}{-q} =
 G_{ee}\frac{q^2K_F^2-({\mathbf kq})^2}
            {2q^2K_F^2\left(q^2+\lambda^2\right)
             -({\mathbf kq})^2\left(2q^2+\lambda^2\right)}
 \ .\label{eq-D11}
\end{equation}

  A set of Eqs.(\ref{eq-D4}), (\ref{eq-D7}), (\ref{eq-D10}), 
and (\ref{eq-D11}) defines the solution of the problem 
when $p,\ k \approx K_F$ and $q \to 0$. 

\end{appendix}
\newpage

\end{document}